\documentclass[aps,pra,reprint,superscriptaddress,floatfix,showpacs]{revtex4-1}
\usepackage{amsmath,bm} 
\usepackage{graphicx}
\usepackage{color}
\usepackage{braket}

\begin{document}

\title{\textit{Ab initio} analysis of the topological phase diagram of the Haldane model}
\author{Julen Iba\~nez-Azpiroz}
\affiliation{\mbox{Peter Gr\"unberg Institute and Institute for Advanced Simulation, Forschungszentrum J\"ulich \& JARA, D-52425 J\"ulich, Germany}}

\author{Asier Eiguren}
\affiliation{Depto. de F\'isica de la Materia Condensada, Universidad del Pais Vasco, UPV/EHU, 48080 Bilbao, Spain}
\affiliation{Donostia International Physics Center (DIPC), 20018 Donostia, Spain}

\author{Aitor Bergara}
\affiliation{Depto. de F\'isica de la Materia Condensada, Universidad del Pais Vasco, UPV/EHU, 48080 Bilbao, Spain}
\affiliation{Donostia International Physics Center (DIPC), 20018 Donostia, Spain}
\affiliation{Centro de F\'{i}sica de Materiales CFM, Centro Mixto CSIC-UPV/EHU, 20018 Donostia, Spain}

\author{Giulio Pettini}
\affiliation{Dipartimento di Fisica e Astronomia, Universit\`a di Firenze,
and INFN, 50019 Sesto Fiorentino, Italy}

\author{Michele Modugno}
\affiliation{\mbox{Depto. de F\'isica Te\'orica e Hist. de la Ciencia, Universidad del Pais Vasco UPV/EHU, 48080 Bilbao, Spain}}
\affiliation{IKERBASQUE, Basque Foundation for Science, 48011 Bilbao, Spain}
\date{\today}


\begin{abstract}
We present an \textit{ab initio} analysis of a continuous Hamiltonian that maps into the celebrated Haldane model. 
The tunnelling coefficients of the tight-binding model are computed by means of two independent methods 
- one based on the maximally localized Wannier functions, the other through analytic expressions in 
terms of gauge-invariant properties of the spectrum - that provide a remarkable agreement and allow 
to accurately reproduce the exact spectrum of the continuous Hamiltonian. By combining these results with the numerical 
calculation of the Chern number, we are able to draw the phase diagram in terms of the physical parameters
of the microscopic model. Remarkably, we find that only a small fraction of the original phase diagram of 
the Haldane model can be accessed, and that the topological insulator phase is suppressed in the deep tight-binding regime.
\end{abstract}

\maketitle

\section{Introduction}
The Haldane model~\cite{haldane} is a celebrated lattice model describing a Chern insulator~\cite{qi2006}, characterized by the presence of quantum Hall effect (QHE)~\cite{klitzing1980} in the absence of a macroscopic magnetic field.
Conceptually, the Haldane model stands at the heart of the tremendous advances in the field of
topological condensed matter physics, as the mechanism for a non-trivial band topology 
presented by Haldane is realized in actual materials via the intrinsic spin-orbit interaction of topological insulators~\cite{hasan2010,qi2011}. 
These concepts are also relevant for the physics of ultracold atoms in optical
lattices, as these systems represent a powerful platform for simulating solid-state physics 
\cite{lewenstein2007}. 
Mott insulators~\cite{greiner2002,schneider2008}, bosonic superfluids~\cite{lim2008} or 
graphene-like honeycomb lattices~\cite{lee2009,lim2014,soltan-panahi2011,soltan-panahi2011b,gail2012,tarruell2012,lim2012} are among the many systems
that have been emulated by this technique. 
Interestingly, an effective experimental realization of the Haldane model has been 
recently reported in Ref. \cite{jotzu2014}.

In his original work, Haldane constructed a discrete tight-binding 
model for a non-centrosymmetric honeycomb lattice in the presence of 
a vector potential $\bm{A}(\bm{r})$, with vanishing total flux through the unit cell. 
The key feature of the model is that, 
even in absence of a macroscopic magnetic field, 
the time-reversal symmetry is broken due to the presence of the gauge field $\bm{A}(\bm{r})$. This, in turn, implies that the next-to-nearest neighbour tunnelling
coefficient $t_{1}$ becomes a complex number.
Haldane showed that the properties of the system
depend on the interplay between the phase acquired by $t_{1}$
and the effect of parity breaking,
affecting the topological phase diagram of
the model~\cite{haldane}. 

Considering the above, the knowledge of the dependence of the  phase acquired by $t_{1}$ 
on the applied vector potential field becomes 
a crucial element for drawing the topological phase diagram.
For this purpose, it is common practice \cite{haldane,shao2008}
to make use of the so-called Peierls substitution, 
whereby the effect of $\bm{A}(\bm{r})$ is effectively included by the replacement 
$t_{1}\rightarrow t_{1}
\exp{(i(e/\hbar)\int \bm{A}(\bm{r}) d\bm{r})}$ \cite{bernevig2013}.
However, in a recent work~\cite{ibanez-azpiroz2014} we showed that the Peierls substitution is actually wrong
whenever the vector field $\bm{A}(\bm{r})$ has the same periodicity of the underlying lattice, 
as it is the case of the Haldane model by construction. 
In that work, we analyzed the parity invariant case by  
presenting two independent approaches for calculating the tight-binding parameters of the model:
one based on the maximally localized Wannier functions (MLWFs), the other
on a closed set of analytical expressions in terms of 
the energy spectrum at selected high symmetry points in the Brillouin zone (BZ).

In the present work, we extend the previous analysis to the general case in which both 
inversion and time-reversal symmetry can be broken.
We show that the two approaches considered provide a remarkable agreement even in the presence of 
parity breaking, allowing for a precise determination of the tight-binding parameters of the model.
By combining these results with the numerical calculation of the Chern number, 
we are able to redraw the topological phase diagram of the Haldane model 
in terms of the physical parameters of the microscopic model. 
Interestingly, we find that only a small fraction of the original phase diagram can be accessed, and that 
the topological insulator phase shrinks dramatically as the system 
becomes more and more tight-binding.
In addition, we find that the gap closing at the topological phase transition does not take place exactly 
at one of high symmetry points of the BZ, but in a close-by point. 
The reason is that the complex tunneling between homologous 
sites are no longer degenerate in the presence of parity breaking, 
contrarily to what it is assumed in the Haldane model.

The paper is organized as follows. In section \ref{sec:method} we introduce the microscopic 
continuous Hamiltonian used in this work and review the formal steps needed to derive the 
corresponding tight-binding model. Some general properties of the Haldane model are also 
recalled. Then, in section \ref{sec:tbparam} we present the two approaches employed for 
calculating the tight-binding parameters, and discuss how the breaking of time-reversal
and/or parity affects their behavior. In section \ref{sec:topo} we 
analyze the topological phase diagram, both in terms of the parameters of the tight-binding model 
and of the physical ones.
Concluding remarks are drawn in section \ref{sec:conclusions}. 
Finally, in the appendices we present an analysis of the spread functional of the MLWFs 
(Appendix \ref{appendix:spread}) and additional remarks on the numerics (Appendix \ref{app:spectrum}). 

\section{Setup of the Haldane model}
\label{sec:method}

In this section we present a systematic derivation of the Haldane model starting from the 
continuous Hamiltonian proposed by Shao \textit{et al.} ~\cite{shao2008} in the context of cold
atoms trapped in optical lattices (see also \cite{ibanez-azpiroz2014}).
The method discussed here is general and suited to map a generic continuous Hamiltonian to 
its corresponding tight-binding model \cite{ibanez-azpiroz2013,ibanez-azpiroz2013b}.

\subsection{The continuous Hamiltonian}
\label{subsec:cont-H}

The general form of a continuous Hamiltonian in the presence of a scalar lattice potential $V_{L}(\bm{r})$ and 
a vector potential $\bm{A}(\bm{r})$ is
\begin{equation}
H_0=\frac{1}{2m}\left[{\bm{p}}-\bm{A}(\bm{r})\right]^{2} + V_{L}(\bm{r}),
\label{eq:h0}
\end{equation}
with $\bm{r}=(x,y)$ in case of a two dimensional system, as we shall consider here. 
The potential $V_{L}(\bm{r})$
is chosen in order to generate a periodic structure with two minima per unit cell,
forming a honeycomb lattice \cite{shao2008,lee2009}:
\begin{eqnarray}
\label{eq:honeycomb}
\begin{split}
V_{L}(\bm{r})=&2sE_R\Big\{\cos\left[(\bm{b}_{1}-\bm{b}_{2})\cdot\bm{r}\right] \\
& + \cos\left(\bm{b}_{1}\cdot\bm{r}-\frac{\pi}{3}\chi_{A}\right)
 +\cos\left(\bm{b}_{2}\cdot\bm{r}\right)\Big\}.
 \end{split}
\end{eqnarray}
Above, $E_{R}=\hbar^{2}k_{L} ^{2}/2m$ is the recoil energy,  $k_L$ denotes the laser wavevector,
 $s$ is a dimensionless parameter representing the strength of the potential in units of $E_{R}$, 
 $\bm{b}_{1,2}=(\sqrt{3}k_{L}/2)(\bm{e}_{x}{\mp}\sqrt{3}\bm{e}_{y})$ are the basis vectors in the reciprocal $\bm{k}$ space, and $\chi_{A}$ is a parameter related to the breaking of the parity symmetry.
In particular, $\chi_{A}=0$ corresponds 
to the inversion symmetric case, where the two minima in the unit cell
are degenerate. 
On the other hand, for $\chi_{A}\neq0$ parity is broken and the minima are no longer degenerate. 
The unit cell (shown in Fig. \ref{fig:tunnelling-arrows})
is generated by the direct lattice vectors 
$\bm{a}_{1,2}=(2\pi/3k_{L})(\bm{e}_{x}{\mp}\sqrt{3}\bm{e}_{y})$.
We also define the lattice vector 
$\bm{a}_{3}=\bm{a}_{1}+\bm{a}_{2}$, that will be useful later on.

We now turn to the vector potential contained in the Hamiltonian (\ref{eq:h0}).
As already mentioned, we employ the form proposed by Shao \textit{et al}~\cite{shao2008}, namely
\begin{equation}
\begin{split}
\bm{A}(\bm{r})=&\alpha\hbar k_{L}\Big[\big(\sin((\bm{b}_{2}-\bm{b}_{1})\cdot\bm{r}) 
+ \\
&\frac12
\sum_{i=1}^{2}(-1)^{i}\sin(\bm{b}_{i}\cdot\bm{r})\Big)\bm{e}_{x} -\frac{\sqrt{3}}{2}\sum_{i=1}^{2}\sin(\bm{b}_{i}\cdot\bm{r})\bm{e}_{y}\Big],
\end{split}
\label{eq:vecaper}
\end{equation}
with $\nabla\cdot\bm{A}(\bm{r})=0$ (Coulomb gauge). The 
flux of the corresponding magnetic field $\bm{B}=\nabla\times \bm{A}$ across the unit cell is null \cite{shao2008}, 
as required for the Haldane model.
In the following analysis, the only variable parameter entering the expression for the vector
potential $\bm{A}(\bm{r})$ is its amplitude $\alpha$.
Notice that for $\alpha=0$ the system is symmetric under time-reversal, whereas this is not the case for $\alpha\neq 0$.

\subsection{The tight-binding model}
\label{subsec:tb-model}

The continuous Hamiltonian (\ref{eq:h0}) can be mapped onto the
tight-binding Haldane model \cite{haldane,shao2008} by following the general procedure discussed in 
\cite{ibanez-azpiroz2013,ibanez-azpiroz2013b,ibanez-azpiroz2014}. The starting point is the (single particle) 
many-body Hamiltonian defined by
\begin{equation}
\hat{\cal{H}}_{0}=\int d\bm{r}~{\hat{\psi}}^\dagger(\bm{r})\hat{H}_{0}{\hat{\psi}}(\bm{r}),
\label{eq:genham}
\end{equation}
with $\hat{\psi}$ a field operator for bosonic or fermionic particles.
Then, when the wells of the lattice potential are deep enough, the field operator
 can be conveniently expanded in terms of a set of functions $w_{\bm{j}\nu}(\bm{r})$ localized at each minimum:
\begin{equation}
\hat{\psi}(\bm{r})\equiv \sum_{\bm{j}\nu}{\hat{a}}_{\bm{j}\nu}w_{\bm{j}\nu}(\bm{r}).
\label{eq:psiexpf}
\end{equation}
Above,  $\nu$ is a band index and ${\hat{a}}^{\dagger}_{\bm{j}\nu}$ (${\hat{a}}_{\bm{j}\nu}$) is
the creation (destruction) operator of a single particle in the $j$th cell,  satisfying the usual commutation (or anti commutation) rules following from those of the field $\hat{\psi}$.

As already mentioned in the introduction,  we shall use the maximally localized Wannier functions (MLWFs)
for composite energy bands  \cite{marzari1997} as basis functions. The MLWFs   
are defined as linear combinations of the 
Bloch eigenstates $\psi_{\nu'\bm{k}}(\bm{r})$,
\begin{equation}
w_{\bm{j}\nu}(\bm{r})=\frac{1}{\sqrt{S_{\cal B}}}
\int_{S_{\cal B}} \!\!d\bm{k} ~e^{-i\bm{k}\cdot\bm{R}_{\bm{j}}}\sum_{\nu'=1}^{N}U_{\nu\nu'}(\bm{k})
\psi_{\nu'\bm{k}}(\bm{r}),
\label{eq:mlwfs}
\end{equation}
where $S_{\cal B}$ represents the volume of the first BZ, and $U\in U(N)$ is a gauge transformation
that is obtained from the 
minimisation of the Marzari-Vanderbilt spread functional, 
$\Omega=\sum_{\nu}\left[\langle \bm{r}^2\rangle_{\nu}-\langle \bm{r}\rangle_{\nu}^{2}\right]$ \cite{marzari1997}. We remark that the presence of the vector potential may significantly affect both the Bloch eigenfunctions $\psi_{\nu'\bm{k}}(\bm{r})$ and the unitary matrices $U_{\nu \nu'}(\bm{k})$ \cite{ibanez-azpiroz2014}. A thorough analysis of the MLWFs is given in section \ref{subsec:mlwf}.

In the following, we shall consider the contribution of the first two Bloch bands only, namely
$\nu,\nu'=1,2$. This is sufficient for constructing the lowest lying MLWFs localized at 
the two lattice sites $A$ and $B$ inside the unit cell.
Then, by considering the following transformation from coordinate to reciprocal space,
$\hat{b}_{\nu{\bm{k}}}=({1}/{\sqrt{S_{\cal  B}}})
\sum_{\bm{j}} ~e^{-i{\bm{k}}\cdot{\bm{R}}_{\bm{j}}}\hat{a}_{\bm{j}{\nu}}$, 
the reduced tight-binding Hamiltonian can be written as 
\begin{equation}
\hat{\cal{H}}_{0}^{tb}\equiv\sum_{\nu\nu'=A,B}\int_{S_{\cal B}} d\bm{k}~h_{\nu\nu'}(\bm{k})
\hat{b}_{\nu\bm{k}}^{\dagger}\hat{b}_{\nu'\bm{k}},
\end{equation}
where the $2\times2$ matrix $h_{\nu\nu'}(\bm{k})$ is given by
\begin{equation}\label{eq:hamh}
\begin{split}
h_{\nu\nu'}(\bm{k})&=\sum_{\bm{j}}e^{i{\bm{k}}\cdot{\bm{R}}_{\bm{j}}}
\langle {w}_{\bm{0}\nu}|\hat{H}_{0}|w_{\bm{j}\nu'}\rangle\\
&=\frac{1}{S_{\cal B}}\sum_{\bm{j}}\int_{S_{\cal B}}d\bm{q}~
e^{i(\bm{k}-\bm{q})\cdot\bm{R}_{\bm{j}}}
\sum_{n}U^{*}_{\nu n}(\bm{q})U_{\nu'n}(\bm{q})\epsilon_{n}(\bm{q}),
\end{split}
\end{equation}
with 
$\bm{R}_{\bm{j}}=\{j_{1}{\bf{a}}_{1}+j_{2}{\bf{a}}_{2}\Big| j_{1},j_{2}=0,\pm 1,\pm 2 \dots\}$
and $\bm{j}$ labels the unit cell.
We remark that
the eigenvalues of $h_{\nu\nu'}(\bm{k})$
coincide with the exact Bloch energies $\epsilon_{n}(\bm{k})$ ($n=1,2$)
if the full expansion 
of neighbouring coefficients $\bm{R}_{\bm{j}}$ is retained.
When the system is in the tight-binding regime ($s\gtrsim5$)~\cite{ibanez-azpiroz2013}, 
it is convenient to truncate the series by retaining
only a finite number of matrix elements
$\langle {w}_{\bm{0}\nu}|\hat{H}_{0}|w_{\bm{j}\nu'}\rangle$, with the eigenvalues of $h_{\nu\nu'}(\bm{k})$
still being a good approximation of the exact energies.
We note that since the functions $w_{\bm{j}\nu}(\bm{r})$
are in general complex (see section \ref{subsec:mlwf}), we may expect the matrix elements
to be complex as well. 

\begin{figure}
\centerline{\includegraphics[width=1.0\columnwidth]{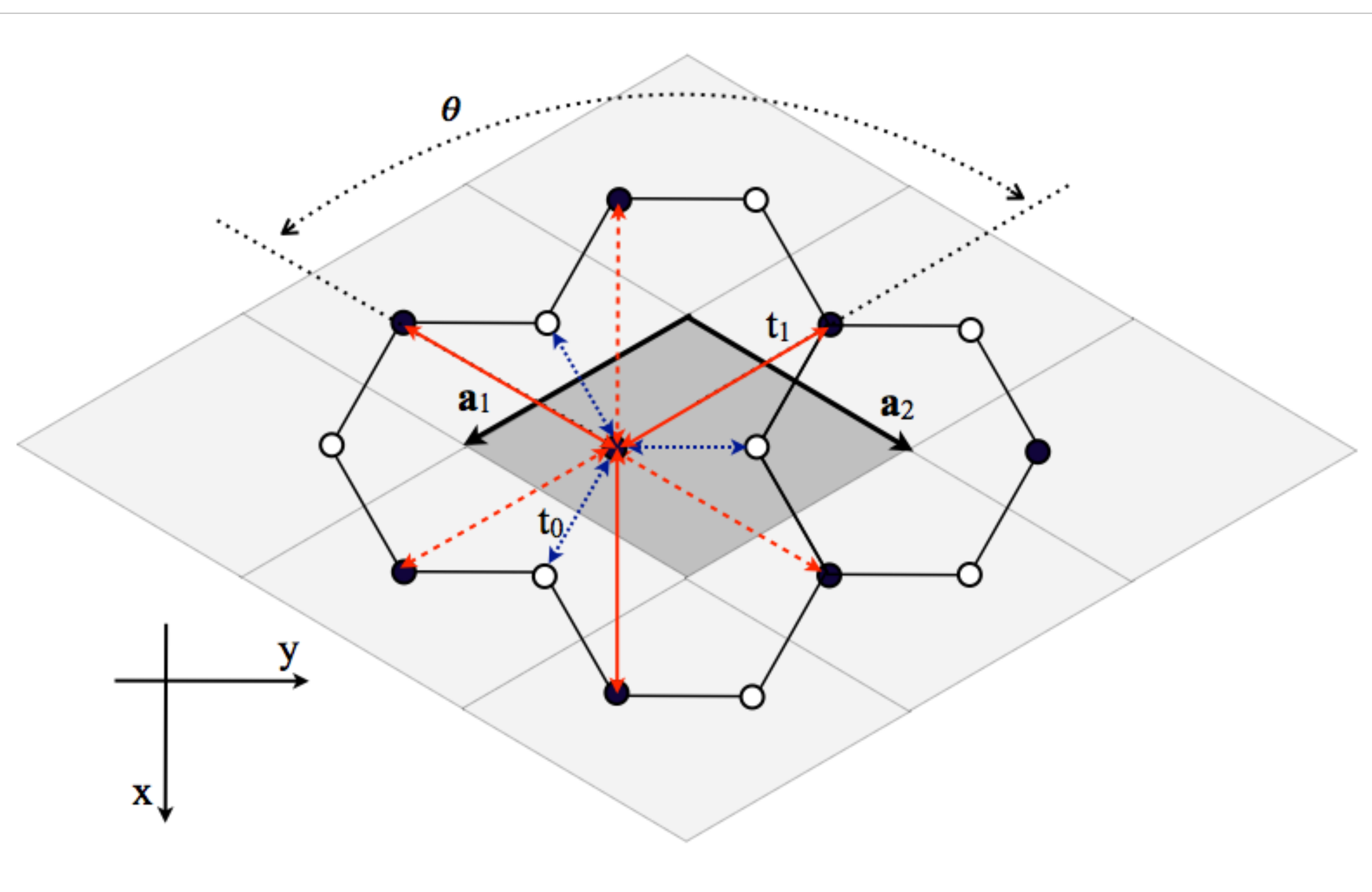}}
\caption{(color online) Bravais lattice associated to the honeycomb potential in  Eq. (\ref{eq:honeycomb}). 
Black and white circles refer to minima of type $A$ and $B$, respectively. 
The elementary cell is highlighted in grey. 
The various tunnelling coefficients are indicated for the site of type $A$ in the central cell. 
The system is invariant under discrete translations generated by the 
Bravais vectors $\bm{a}_{1/2}$ and under rotations of 
$\theta=2\pi/3$ radians around any vertex of the lattice.
The rotational symmetry implies that next-to-nearest tunnelling amplitudes 
$t_{1}$ along the same direction are conjugate pairs (solid and dashed lines in red); 
from the latter follows the equivalence of the hopping amplitudes separated by 
$2\pi/3$ radians. 
When sites $A$ and $B$ are degenerate, 
the system is also invariant 
under rotations by $\pi$ radians around the centre of any elementary cell.
}
\label{fig:tunnelling-arrows}
\end{figure}

By truncating the tight-binding expansion in Eq. (\ref{eq:hamh}) 
to the next-to-nearest neighbours (see Fig. \ref{fig:tunnelling-arrows}), the matrix $h_{\nu\nu'}(\bm{k})$ can be written as the sum of three terms:
\begin{equation}
\label{eq:tb-truncated}
h_{\nu\nu'}(\bm{k})= [h^{(0)}_{\nu\nu'}(\bm{k})+
h^{(2)}_{\nu\nu'}(\bm{k})]\delta_{\nu\nu'}+ h^{(1)}_{\nu\nu'}(\bm{k}).
\end{equation}
The first term corresponds to the on-site energies
\begin{equation}
h^{(0)}_{\nu\nu}(\bm{k})\equiv E_{\nu}=\langle {w}_{\bm{0}\nu}|\hat{H}_{0}|w_{\bm{0}\nu}\rangle,
\end{equation}
which are real quantities by definition. The next term, $h^{(1)}_{\nu\nu'}$, contains 
only off-diagonal elements  
corresponding to the hopping between the three nearest-neighbour sites. 
Although the basis functions $w_{\bm{j}\nu}(\bm{r})$ are complex,
these three tunnelling amplitudes can be chosen to be real by means of a suitable global gauge fixing, as they are all equal thanks to the symmetries of the system (see Fig. \ref{fig:tunnelling-arrows}). 
Taking this into consideration, and defining 
\begin{equation}
\label{eq:t0}
t_{0}\equiv\langle {w}_{\bm{0}A}|\hat{H}_{0}|w_{\bm{0}B}\rangle,
\end{equation}
we can write
\begin{equation}
\label{eq:h1AB}
h^{(1)}_{AB}(\bm{k})=t_{0}\left(1+e^{i\bm{k}\cdot\bm{a}_{1}}+
e^{-i\bm{k}\cdot\bm{a}_{2}}\right)\equiv t_{0}Z_{0}(\bm{k})\equiv z(\bm{k}).
\end{equation}
Its conjugate counterpart is given by $h^{(1)}_{BA}(\bm{k})= z^{*}(\bm{k})$.
Finally, the term
$h^{(2)}_{\nu\nu}(\bm{k})$ 
corresponds to the six next-to-nearest neighbours hopping between homologous sites.
By taking into account all the symmetries of the full Hamiltonian of Eq. (\ref{eq:h0}), 
these tunnelling coefficients can be compactly written as
\begin{equation}
\label{eq:t1}
t^{\pm}_{1\nu}=\langle {w}_{\bm{0}\nu}|\hat{H}_{0}|w_{\pm\bm{a_{j}}\nu}\rangle\equiv|t_{1\nu}|e^{\pm i\varphi_{\nu}},
\;\; j = 1,2,3.
\end{equation}
The above equation explicitly shows that $t^{\pm}_{1\nu}$ contains two distinct 
complex phases 
$\pm\varphi_{\nu}$ for each site type ($\nu=A,B$). 
Then, using Eq. (\ref{eq:t1}) and after some algebra, we can write
\begin{equation}
\label{eq:h2}
\begin{split}
h^{(2)}_{\nu\nu}(\bm{k})=&|t_{1\nu}|\Big\{2\cos\left[\bm{k}\!\cdot\!\bm{a}_{3}
+\varphi_{\nu}\right]+
2\sum_{i=1,2}\cos\left(\bm{k}\!\cdot\!\bm{a}_{i}-\varphi_{\nu}\right)\Big\}\\
&\equiv |t_{1\nu}|F_{\nu}(\bm{k})\equiv f_{\nu}(\bm{k}).
\end{split}
\end{equation}

The above expressions allow to cast the matrix $h_{\nu\nu'}(\bm{k})$ in the following compact form,
\begin{equation}
h_{\nu\nu'}(\bm{k})=\left(\begin{array}{cc}
 \epsilon_{A}(\bm{k}) & z(\bm{k}) \\
 z^{*}(\bm{k}) & \epsilon_{B}(\bm{k})
\end{array}\right),
\label{eq:hmatrix}
\end{equation}
where we have defined
\begin{equation}
\epsilon_{\nu}(\bm{k})= E_{\nu}+f_{\nu}(\bm{k}).
\end{equation}

By expanding the Hamiltonian on the basis of the $2\times 2$ identity matrix, $I$,  
and of the Pauli matrices, $\sigma_{i}$,
Eq. (\ref{eq:hmatrix}) can be rewritten as \cite{shao2008}
\begin{equation}
 h(\bm{k})=h_{0}(\bm{k})I+\sum_{i=1}^{3}h_{i}(\bm{k})\sigma_{i},
\end{equation}
where the coefficients $h_{i}(\bm{k})$ are given by the following expressions:
\begin{eqnarray}
h_{0}(\bm{k})&=&\frac{\epsilon_{A}(\bm{k})+\epsilon_{B}(\bm{k})}{2}=\frac{f_{A}
(\bm{k})+f_{B}(\bm{k})}{2}
\label{generalhmatrixh0}\\
&\equiv& f_{+}(\bm{k}),\notag\\
h_{1}(\bm{k})&=& {\rm{Re}}[z(\bm{k})]=t_{0}\sum_{i=1}^{3}\cos\left(\bm{k}\cdot\bm{s}_{i}\right),\label{generalhmatrixh1}\\
h_{2}(\bm{k})&=& -{\rm{Im}}[z(\bm{k})]= t_{0}\sum_{i=1}^{3}\sin\left(\bm{k}\cdot\bm{s}_{i}\right),
\label{generalhmatrixh2}\\
h_{3}(\bm{k})&=&\frac{\epsilon_{A}(\bm{k})-\epsilon_{B}(\bm{k})}{2}=\epsilon+
\frac{f_{A}(\bm{k})-f_{B}(\bm{k})}{2}\label{generalhmatrixh3}\\
&\equiv&  \epsilon + f_{-}(\bm{k})\notag,
\end{eqnarray}
with the vectors ${\bm{s}}_{i}$ being those  
joining the three nearest neighbours of type $A(B)$ to a given site of type
$B (A)$ \cite{shao2008}. In the last expression we have also fixed, without loss of
generality, $E_{A}=\epsilon$ and $E_{B}=-\epsilon$.

Then, the tight-binding energies are readily found as 
\begin{equation}
\epsilon_{\pm}(\bm{k})= h_{0}(\bm{k})\pm|{\bm{h}}(\bm{k})|
= f_{+}(\bm{k})\pm\sqrt{[\epsilon+f_{-}(\bm{k})]^2+|z(\bm{k})|^2},
\label{eq:tbenergies}
\end{equation}
where ${\bm{h}}\equiv(h_{1},h_{2},h_{3})$.

\subsubsection{The Haldane model and the Peierls substitution.}
\label{subsubsec:SPS}
At this point, further approximations are required in order to recover 
the original model proposed by Haldane \cite{haldane}, namely 
$|t_{1A}|=|t_{1B}|\equiv|t_{1}|$ and $\varphi_A=-\varphi_B\equiv\varphi$. 
We shall refer to this configuration as the ``simplified parameter setup'' (SPS). We note that the corresponding model
contains only four parameters, namely $\epsilon, t_0,|t_{1}|$ and $\varphi$.
In section \ref{sec:results}, we shall provide numerical evidence showing that
in the tight-binding regime the difference between $|t_{1A}|$ and $|t_{1B}|$ is
negligible, thus justifying the SPS. 
The second condition is not strictly verified (again, see section \ref{sec:results}),
but one can consider a sort of effective model by defining a single phase,
$\varphi\equiv(\varphi_A-\varphi_B)/2$. Therefore, 
in the SPS the terms 
$h_0$ and $h_3$ of Eqs. (\ref{generalhmatrixh0}) and (\ref{generalhmatrixh3}) simplify to  
\begin{eqnarray}
h_{0}&=&2|t_{1}|\cos\varphi\sum_{i=1}^{3}
\sin\left(\bm{k}\cdot\bm{a}_{i}\right),\nonumber\\
h_{3}&=&\epsilon-2|t_{1}|\sin\varphi\sum_{i=1}^{3}
\sin\left(\bm{k}\cdot\bm{a}_{i}\right).
\label{h12hald}
\end{eqnarray}
The above equations, together with Eqs. (\ref{generalhmatrixh1}) and (\ref{generalhmatrixh2}), 
correspond to the definition of the original Haldane model, see \cite{haldane,shao2008}.

As already mentioned in the introduction, the original model proposed by Haldane  
is constructed by means of the so-called Peierls substitution~\cite{haldane,shao2008}.
This consists in assuming that the complex phase of the tunnelling coefficient $t_{ij}$ is given by 
the integral of the vector potential along the straight path joining sites $i$ and $j$, 
\textit{i.e.} $t_{ij}\to t_{ij}\exp\{i(e/\hbar)\int_{i}^{j}\bm{A}d\bm{r}\}$. 
In the present case, the Peierls prediction for the complex phase would be~\cite{shao2008}
\begin{equation}
\label{eq:peierls-phase}
\varphi_{P}=\frac{2\pi}{\sqrt{3}}\alpha.
\end{equation}
This value will be used later on, in Secs. \ref{sec:tbparam} and \ref{sec:topo},
for comparison with the results of the two approaches discussed in this paper. 
Here, we can anticipate - as thoroughly discussed in \cite{ibanez-azpiroz2014} - that the above result is definitely incorrect, owing to the fact that the Peierls substitution is justified only when the vector field $\bm{A}(\bm{r})$ is slowly varying on the scale of the lattice~\cite{boykin2001}. In fact, this condition is explicitly violated in the Haldane model, where $\bm{A}(\bm{r})$ has the same periodicity of the lattice.

\subsection{General features of the Haldane model}
\label{sec:haldane}

Before proceeding, let us recall some general features of the Haldane model \cite{haldane}, corresponding to the SPS approximation.
This model is characterized by the presence of Dirac points located at the vertices 
$\bm{k}_D$ of the first BZ, where the  dispersion law takes the relativistic form
$\epsilon(\bm{k})=\sqrt{m^{2}c^{4}+c^{2}k^{2}}$. They can be divided into two inequivalent 
classes, corresponding e.g. to $\bm{k^{\pm}}_D=\pm(1,0){k_{L}}$ (often referred in the 
literature as K and K$^{\prime}$; in the following, we will always 
refer to these two as inequivalent Dirac points, for simplicity).
In the presence of time-reversal and inversion symmetry (namely, for $\alpha=0$, $\chi_A=0$), 
the two energy bands become degenerate at the Dirac points, whose position is given by the solution of $z(\bm{k}_{D})=0$
or $h_1(\bm{k}_D)=h_2(\bm{k}_D)=0$ (this holds at any order of the tight-binding expansion).
In fact, in this case both $\epsilon$ and $\varphi$ vanish, yielding $h_3(\bm{k})=0$, $f_A=f_B$, 
and $f_{-}=0$, so that the energies $\epsilon_{\pm}(\bm{k}_D)$ are degenerate.

When both time-reversal and inversion symmetry are broken ($\alpha\neq 0,\chi_A\neq 0$), two inequivalent energy gaps 
form at the Dirac points:
\begin{equation}
 \delta_{\pm}\equiv \epsilon_{+}(\bm{k}^{\pm}_{D})-\epsilon_{-}
 (\bm{k}^{\pm}_{D})= 2|h_{3}(\bm{k}^{\pm}_{D})|=2\sqrt{[\epsilon+f_{-}(\bm{k}^{\pm}_{D})]^2}.
\end{equation}
The closure of  one of them indicates a topological phase transition, where
\begin{equation}
\delta_{\pm}\equiv2\left|\epsilon\pm 3\sqrt{3}|t_1|\sin\varphi\right|=0.
\label{eq:gaphaldane}
\end{equation}
This equation identifies  
the well known boundary between the normal and topological insulator phases 
with Chern numbers $C=0$ and $C=\pm 1$ in the Haldane model \cite{haldane}.  

We remark that in the general model the gap closing 
does not take place exactly at $\bm{k}^{\pm}_{D}$, but in a close-by point. 
This is due to the fact that, when breaking of parity is included self consistently,
the tunnelling parameters $t_{1A}$ and $t_{1B}$ are no longer degenerate, contrarily to what it is assumed in the Haldane model (cf. the SPS approximation).

\section{Calculation of the tight-binding parameters}
\label{sec:tbparam}

In this section we discuss two independent methods for calculating the 
tight-binding parameters for arbitrary values of the physical parameters 
$s,\alpha$ and $\chi_A$. The first method is based on the \textit{ab initio} 
calculation of the maximally localized Wannier functions (MLWFs) \cite{marzari1997,marzari2012},  
which we already employed in \cite{ibanez-azpiroz2014,ibanez-azpiroz2013,ibanez-azpiroz2013b} in different lattice geometries. 
This approach gives direct access to the whole set of parameters 
$\epsilon, t_0,|t_{1A}|,|t_{1B}|,\varphi_A$ and $\varphi_B$. The second approach relies instead on analytical expressions in terms of the energy spectrum,
as discussed in \cite{ibanez-azpiroz2014}. The latter is here extended to the case of parity breaking.

We remark that the approach based on the MLWFs corresponds to the \textit{ab initio} definition of the parameters, and is therefore model-independent.
Instead, the second method depends on the specific form of the tight-binding Hamiltonian. However, it does not require the calculation of any set of Wannier functions since only the spectrum of the continuous Hamiltonian is needed.
Notably, the two methods present a remarkable agreement in the whole range of parameters considered here.

\subsection{Maximally localized Wannier functions}
\label{subsec:mlwf}

The MLWFs, which have been defined in Eq. (\ref{eq:mlwfs}), represent  a powerful tool that is largely 
employed in condensed matter physics~\cite{marzari2012}.
By construction, the MLWFs are the basis functions with the maximal degree of localization in coordinate space, allowing to construct tight-binding models that accurately reproduce the properties of the continuous Hamiltonian, with a minimal set of tunnelling coefficients.
In addition, the MLWFs permit a very fine sampling of the reciprocal 
space thanks to the so-called Wannier interpolation technique \cite{marzari2012}.
This point is very important for our purposes in this work, as 
the determination of the Chern number 
requires a high density of points in $\bm{k}$-space~\cite{wang2007,wang2006}.

The MLWFs  are computed by means of the standard implementation of 
the WANNIER90 package~\cite{mostofi2008} (see also Appendix \ref{app:spectrum}). The resulting
functions are complex-valued when $\alpha\neq0$. 
This feature is in agreement with the analysis of 
 \cite{brouder2007}, where it was shown 
that, in general, MLWFs cannot be constructed as real functions when the time-reversal symmetry is broken
(see also \cite{panati2013}).
In the context of the Haldane model, the imaginary part of the MLWFs plays an
essential role, since it determines the 
complex phase acquired by the next-to-nearest tunnelling coefficient. 
In turn, the complex phase directly affects 
physically meaningful quantities, such as the spectrum or
the topological phase diagram~\cite{haldane}. 

\begin{figure}
\centerline{\includegraphics[width=0.7\columnwidth]{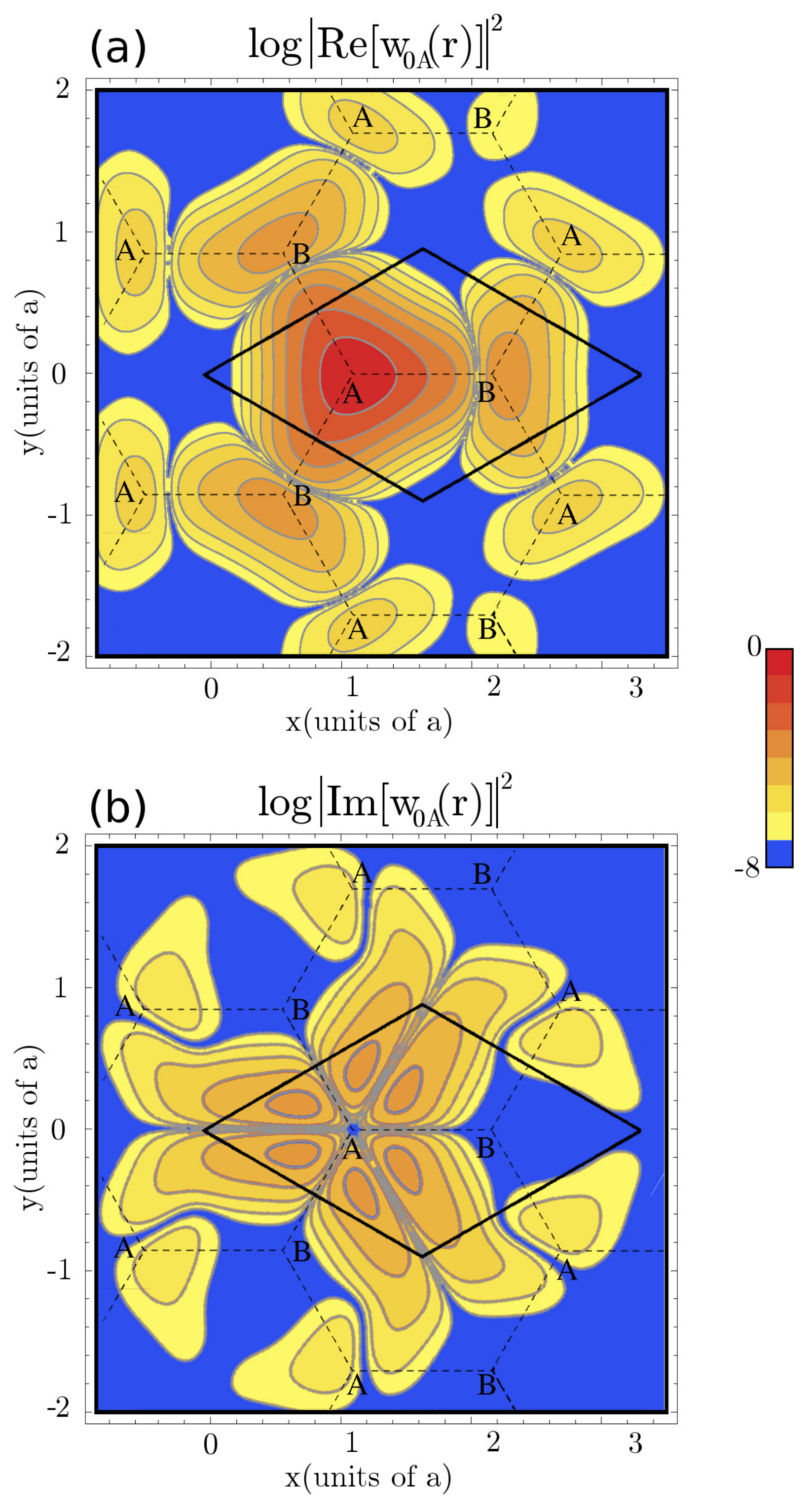}}
\caption{(Color online) Density plot (in logarithmic scale) of the 
square of the real (a) and imaginary (b) parts of the 
MLWF for sublattice A,  for  $s=5$, $\alpha=0.1$ and $\chi_{A}=0$. 
The solid and dashed lines denote the unit cell and
the honeycomb lattice of the scalar potential, respectively. 
In the latter, the corners of the hexagons mark the minima of the 
scalar lattice potential labelled either as $A$ or $B$. }
\label{fig:im-mlwf}
\end{figure}

In Fig. \ref{fig:im-mlwf} we illustrate an example of 
the real-space structure (note the logarithmic scale) 
of the real and imaginary parts
of a MLWF for sublattice $\nu=A$, located at the origin $\bm{j}=\bm{0}$, 
for $s=5$, $\alpha=0.1$ and $\chi_{A}=0$.
The structure of the real part is very similar to  
the one in the pure honeycomb lattice~\cite{ibanez-azpiroz2013},
namely it is highly localized around the origin, with appreciable contribution
around the neighbouring lattice sites.
In average,   
the imaginary part is two-three orders of magnitude smaller than the real part.
It is particularly interesting to observe that the imaginary part 
is null at the interstitial region between nearest neighbours, 
while it
becomes maximum along the path joining next-to-nearest neighbours.
These properties hold in the whole range of parameters considered in this work.

An analysis of the spread functional of the MLWFs as a function of the vector potential amplitude 
has been included in Appendix \ref{appendix:spread}.

\subsection{Analytical expressions from the spectrum}
\label{subsec:assumptions-haldane}

In this section we derive a closed set of analytical expressions in terms of 
the energy spectrum at selected high symmetry points in the BZ.
This is done in the framework of the SPS  discussed in section \ref{subsubsec:SPS},
corresponding to the standard formulation of the Haldane model \cite{haldane,shao2008}.
As we shall see below, the approximations of the SPS are 
well justified in the tight-binding regime.
The model is therefore given in terms of four parameters, namely 
$\epsilon,\varphi,t_0$ and $|t_1|$.
We remind that $\epsilon$ measures the difference between the on-site
energies $E_{A}$ and $E_{B}$, and it is therefore associated to the breaking of parity, 
whereas the breaking of the time-reversal symmetry corresponds  to $\varphi$ different 
from zero. It is also worth recalling that the parameters of the underlying continuous 
Hamiltonian that control the breaking of parity and time-reversal symmetry are $\chi_A$ and
$\alpha$, respectively. In particular, $\chi_A=0$ gives $\epsilon=0$ whereas $\alpha=0$ implies $\varphi=0$. 

We begin by noting the following relations at $\bm{k}=\bm{0}$:
\begin{eqnarray}
f_+(\bm{0})&=&6|t_1|\cos\varphi,\\
f_-(\bm{0})&=&0,\\
|z(\bm{0})|&=&3|t_0|.
\end{eqnarray}
Similarly, at the Dirac points $\bm{k^{\pm}}_D$ we have
\begin{eqnarray}
f_+(\bm{k}^{\pm}_D)&=&-3|t_1|\cos\varphi,\\
f_-(\bm{k}^{\pm}_D)&=&\pm 3\sqrt{3}|t_1|\sin\varphi,\\
z(\bm{k}^{\pm}_D)&=&0.
\end{eqnarray}
Next, let us  define the bandwidths 
\begin{eqnarray}
\Delta^{\pm}_{+}&=&+[\epsilon_{+}(\bm{0})-\epsilon_{+}(\bm{k}^{\pm}_D)],\\
\Delta^{\pm}_{-}&=&-[\epsilon_{-}(\bm{0})-\epsilon_{-}(\bm{k}^{\pm}_D)].
\label{eq:bandwidths}
\end{eqnarray}
Recalling the expression for the gap at the Dirac points in Eq. (\ref{eq:gaphaldane}),
one can easily derive the following relations:
\begin{eqnarray}
\label{eq:11}
\sqrt{\epsilon^2+9t_0^2}&=&\frac{\Delta^{+}_++\Delta_-^++\delta_{+}}{2}=\frac{\Delta^{-}_++\Delta_-^-+\delta_{-}}{2},\\
\label{eq:22}
18|t_1|\cos\varphi&=&\Delta^{+}_+-\Delta_-^+=\Delta^{-}_+-\Delta_-^- .
\end{eqnarray}
Due to the symmetries of the system, 
we can consider $\epsilon\geq 0$ and $\varphi\geq 0$ without loss of generality. 
Focusing first on the region with $\epsilon > 3\sqrt{3}|t_{1}|\sin\varphi$ (corresponding to the normal insulator phase), after some algebra one finds the following set of formulas:
\begin{eqnarray}
\label{eq:eps-spc}
\epsilon&=&\frac{\delta_{+}+\delta_{-}}{4},\\
\label{eq:t0-spc}
t_{0}&=&\frac{1}{6}\sqrt{\left(\Delta_+^++\Delta_-^++\delta_{+}\right)^2-\frac{\left(\delta_{+}+\delta_{-}\right)^2}{4}},
\\
\label{eq:t1-spc}
|t_{1}|&=&\frac{1}{18}\sqrt{\left(\Delta_+^+-\Delta_-^+\right)^2+\frac{3}{4}\left(\delta_{+}-\delta_{-}\right)^2},\\
\label{eq:phi-spc}
\varphi&=&\rm{tg}^{-1}\left[\frac{\sqrt{3}}{2}\frac{\delta_{+}-\delta_{-}}{\Delta_+^+-\Delta_-^+}\right].
\end{eqnarray}
Similarly, in the region with $\epsilon < 3\sqrt{3}|t_{1}|\sin\varphi$ (corresponding to the topological insulator phase), we find the following expressions:
\begin{eqnarray}
\label{eq:eps-spc1}
\epsilon&=&\frac{\delta_{+}-\delta_{-}}{4},\\
\label{eq:t0-spc1}
t_{0}&=&\frac{1}{6}\sqrt{\left(\Delta_+^++\Delta_-^++\delta_{+}\right)^2-\frac{\left(\delta_{+}-\delta_{-}\right)^2}{4}},
\\
\label{eq:t1-spc1}
|t_{1}|&=&\frac{1}{18}\sqrt{\left(\Delta_+^+-\Delta_-^+\right)^2+\frac{3}{4}\left(\delta_{+}+\delta_{-}\right)^2},\\
\label{eq:phi-spc1}
\varphi&=&\rm{tg}^{-1}\left[\frac{\sqrt{3}}{2}\frac{\delta_{+}+\delta_{-}}{\Delta_+^+-\Delta_-^+}\right].
\end{eqnarray}
The solutions in a generic case with $\epsilon<0$ or $\varphi<0$ can be obtained from symmetry considerations,
by exchanging the role of the two basis points $A,B$ and/or of the two inequivalent Dirac 
points $\bm{k}^{\pm}_D$.

\subsection{Numerical results}
\label{sec:results}

In this section we present a comparison of the two methods described 
in Secs. \ref{subsec:mlwf} and \ref{subsec:assumptions-haldane}
for the calculation of the tight-binding parameters.
In addition, we also analyze the accuracy of the assumptions of the SPS
(Sec. \ref{subsec:assumptions-haldane}) based on the tunneling coefficients 
extracted from the MLWFs.
\begin{figure}[t]
\centerline{\includegraphics[width=0.7\columnwidth]{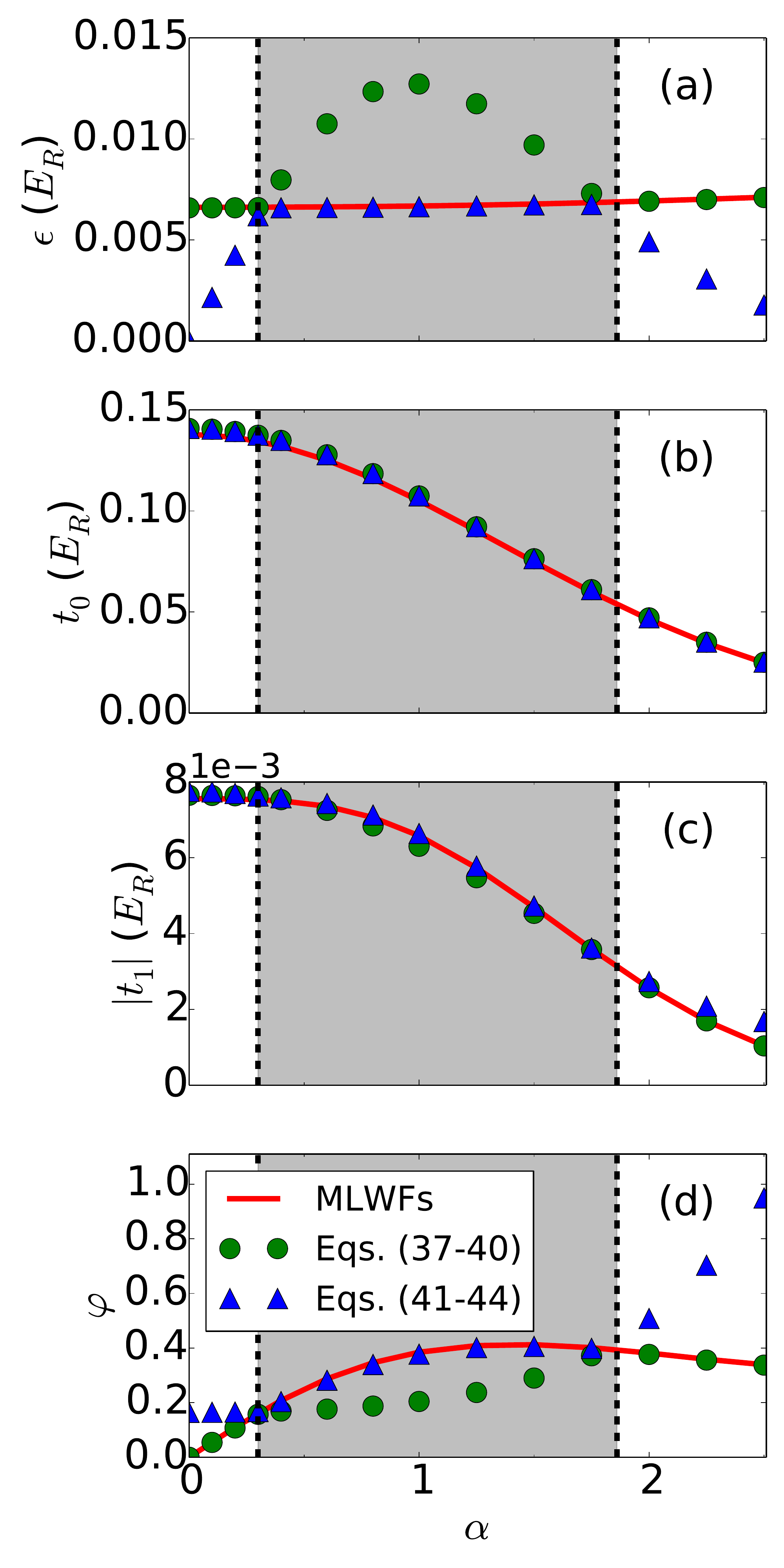}}
\caption{(color online) Comparison of the four tight-binding coefficients 
as calculated from the MLWFs (solid red lines) 
and the analytical formulas of Eqs. (\ref{eq:eps-spc})-(\ref{eq:phi-spc})
(blue triangles) and Eqs. (\ref{eq:eps-spc1})-(\ref{eq:phi-spc1})
(green circles).
Results are shown as a function of $\alpha$,  keeping fixed values $s=5$
and $\chi_{A}=0.001$. 
The grey area in the figures denotes the region where 
the system behaves as a topological insulator with $C\neq 0$ (see text and section \ref{sec:topo}). 
}
\label{fig:t0-t1-phi}
\end{figure}

Let us begin by analyzing Fig. \ref{fig:t0-t1-phi}, where  we compare the tunnelling coefficients 
calculated from the MLWFs with those calculated from the analytical formulas
of Eqs. (\ref{eq:eps-spc})-(\ref{eq:phi-spc}), valid for 
the normal insulator regime, and Eqs. (\ref{eq:eps-spc1})-(\ref{eq:phi-spc1}),
valid for the topological insulator regime, which is depicted by the grey shaded area in the figure.
Results  are shown
as a function of $\alpha$ for fixed values $s=5$ and $\chi_{A}=0.001$, 
since  essential features are unaffected by $s$ and $\chi_{A}$.
In the case of the MLWFs, we have plotted the averages $\varphi=(\varphi_A-\varphi_B)/2$ 
and $|t_{1}|=(|t_{1A}|+|t_{1B}|)/2$ in order to allow comparison with 
the analytical formulas, 
which have been derived in the context of the SPS (see section \ref{subsec:assumptions-haldane})

Overall, Fig. \ref{fig:t0-t1-phi} shows a very good agreement between the two methods
for all the tunnelling coefficients, in all regimes.
Furthermore, it is interesting to note that the two different solutions 
represented by the set of Eqs. (\ref{eq:eps-spc})-(\ref{eq:phi-spc}) 
and (\ref{eq:eps-spc1})-(\ref{eq:phi-spc1}) exchange roles at the boundaries
between normal and topological insulator regimes; this feature is particularly noticeable
in Figs. \ref{fig:t0-t1-phi}(a) and \ref{fig:t0-t1-phi}(d).
To put in other words, the solution of one set of equations on one side
represents a smooth continuation of the solution of the other set of equations in the other side,
and viceversa. Provided that one chooses the right solution, the calculated values agree
very well with those of the MLWFs, as already said.
In addition, 
Fig. \ref{fig:t0-t1-phi}(d) reveals an extremely important feature that was absent in the original Haldane model: 
the phase $\varphi$ is limited by a maximal value. 
This behavior, that was already found in the parity-symmetric case \cite{ibanez-azpiroz2014}, implies that $\varphi$ can only access a restricted range of values, therefore
limiting the physically accessible region of the phase diagram. This feature 
will be crucial for the analysis presented in the next section, where we shall redraw the topological phase 
diagram in terms of the physical parameters -
$\alpha,\chi_A$ and $s$ - of the underlying continuous Hamiltonian.

\begin{figure}
\centerline{\includegraphics[width=0.55\columnwidth]{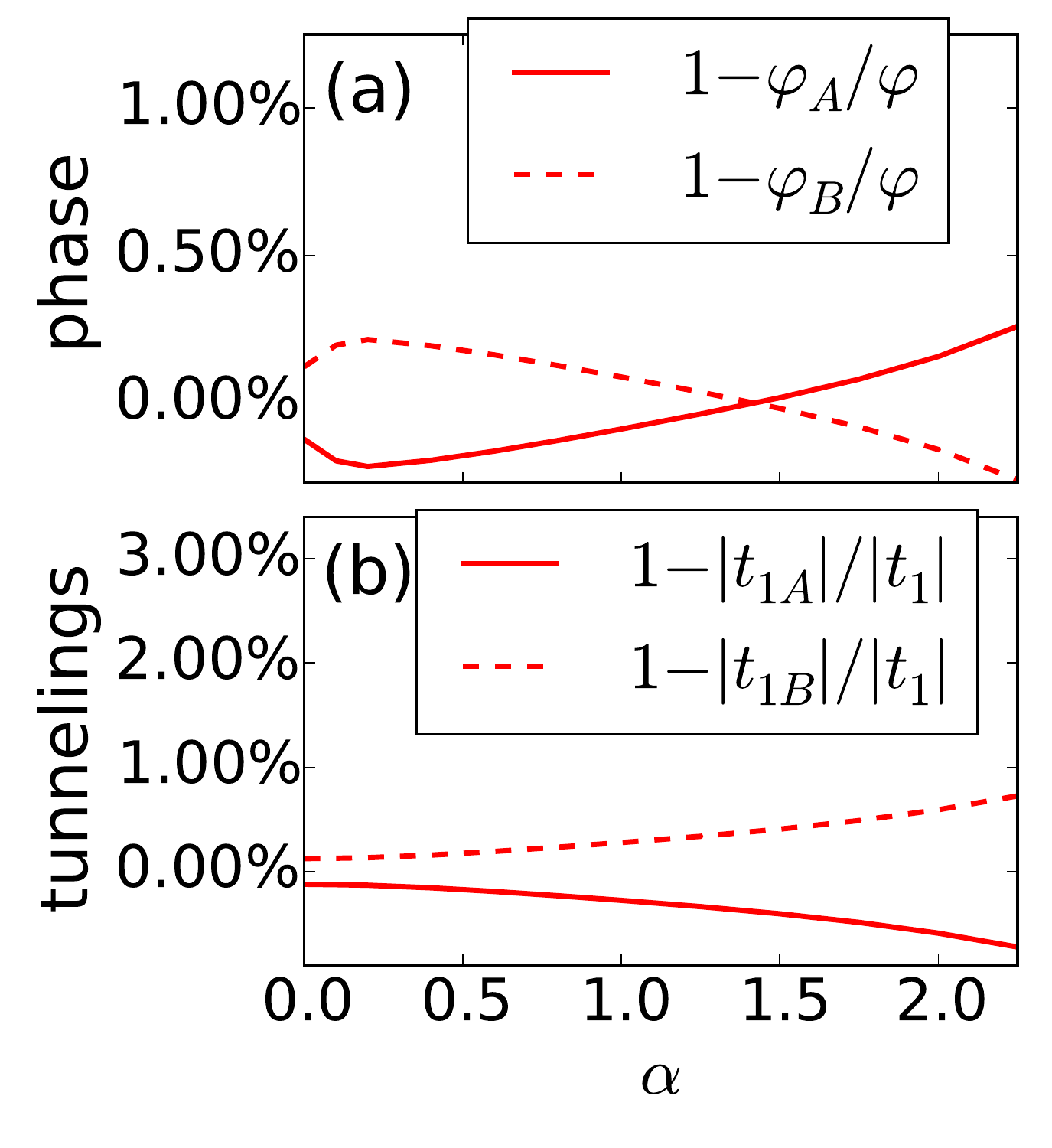}}
\caption{(color online) Relative deviations from the average values of (a) 
the phase, $1-\varphi_{A,B}/\varphi$, and (b) the magnitude of the next-to-nearest tunneling coefficient, 
$1-|t_{1A,B}|/|t_{1}|$,  for  $\chi_{A}=0.001$, $s=5$ $E_{R}$.
Results calculated using the MLWFs.}
\label{fig:sps}
\end{figure} 

Next, we proceed to test the accuracy  of the 
assumptions of the SPS approximation (Sec. \ref{subsec:assumptions-haldane}) based on the tunneling coefficients
calculated from the MLWFs. 
This is done in Fig. \ref{fig:sps}, where we compare the relative deviations from the average values of the phase, $1-\varphi_{A,B}/\varphi$, and of the magnitude of the next-to-nearest tunneling coefficient, 
$1-|t_{1A,B}|/|t_{1}|$, for $\chi_{A}=0.001$, $s=5$. 
This figure demonstrates that the maximum relative deviation in both cases is
below $\sim 1\%$. 
We have verified that this holds for all values of $s$ and $\chi_{A}$ considered here, 
thus justifying the assumptions of the SPS approximation
in the whole range of parameters.
Apart from the relative deviation, Fig. \ref{fig:sps}(a) reveals that $\varphi_{A}$ and $\varphi_{B}$
exchange roles at $\alpha\sim1.5$, around the point
where the phase gets its maximum value, see Fig. \ref{fig:t0-t1-phi}(d).

\section{Topological phase diagram}
\label{sec:topo}

The topological state of a system is 
characterized by the so-called Chern number or topological index~\cite{thouless1982}
\begin{equation}
\label{eq:chern}
C=\frac{i}{2\pi}\int_{BZ}\!\!\!\!d\bm{k}
\sum_{\nu}^{occ}
\braket{\partial_{\bm{k}}u_{\nu\bm{k}}|\times|\partial_{\bm{k}}u_{\nu\bm{k}}},
\end{equation}
with $u_{\nu\bm{k}}(\bm{r})=e^{-i\bm{k}\cdot\bm{r}}\psi_{\nu\bm{k}}(\bm{r})$
being the periodic part of the Bloch eigenfunctions. 
Since the band structure of the Haldane model 
consists on a valence and a conduction band, 
only the lower energy band enters the sum over occupied states in  Eq. (\ref{eq:chern}).  
In order to efficiently calculate the Chern number, one can 
rewrite the expression in (\ref{eq:chern}) as
\begin{equation}
\label{eq:chern-berry}
C=\frac{1}{2\pi}\int_{BZ}\!\!\!\!\!\!d\bm{k}~\Omega(\bm{k}),
\end{equation}
where $\Omega(\bm{k})$ stands for the Berry curvature~\cite{fujita2011}.
This quantity can be accurately computed by means of the Wannier interpolation technique, as discussed
in \cite{wang2006,lopez2012,mostofi2008}.  
In our calculations, we find that a fine $5000\times5000$ $\bm{k}$-mesh is required in order to 
converge the integral of Eq. (\ref{eq:chern-berry}).  

The Chern number represents a topological property and takes 
only integer numbers~\cite{thouless1982}. 
Its value is intimately connected to the band structure
and the gaps opened by symmetry breaking at the Dirac points. 
If a gap is opened solely by inversion symmetry breaking, 
the state of the system is topologically trivial with $C=0$.
On the other hand, if the gap is opened by time-reversal symmetry breaking, then
the system is found in a topologically non-trivial state with $C\neq 0$.
When both symmetries are broken, the topological state of the system 
depends on the relative strength of the inversion and time-reversal symmetry breaking. 

\begin{figure}[t]
\centerline{\includegraphics[width=1.00\columnwidth]{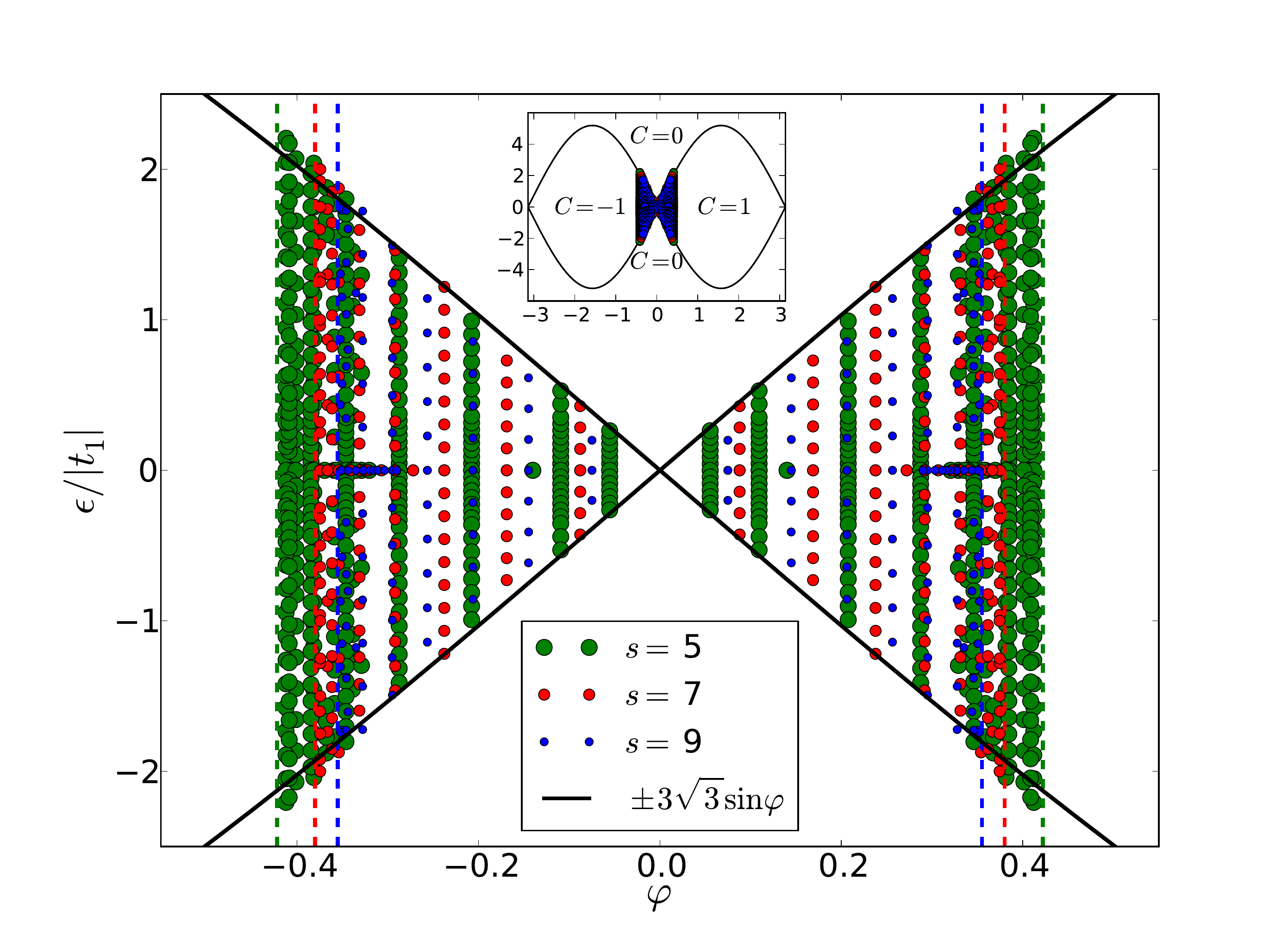}}
\caption{(Color online) Topological phase
diagram of the Haldane model 
as a function of $\varphi$ and $\epsilon/|t_{1}|$.
The main figure is a zoom for  $\varphi\in [-0.45,0.45]$,
while the inset illustrates the full nominal diagram, with $\varphi\in [-\pi,\pi]$. 
Large (green), medium-size (red) and small (blue) 
dots correspond to non-zero Chern numbers
calculated \textit{ab initio} for $s=5,7$ and $9$, respectively. 
The sign of the Chern number is equal to the sign of the phase.
The solid (black) line denotes the
analytical boundary $\epsilon/|t_{1}|=3\sqrt{3}\sin\varphi$.
The vertical dashed lines delimit the physically accessible regions. 
}
\label{fig:C-phi-M-t1}
\end{figure}

The topological phase diagram of the Haldane model has been traditionally 
drawn as a function of $\varphi$ and $\epsilon/|t_{1}|$~\cite{haldane,shao2008}. 
In order to facilitate the discussion, let us rewrite here 
the analytic expression in Eq. (\ref{eq:gaphaldane})
that defines the boundary between the different insulating regions, namely 
\begin{equation}
\label{eq:gaphaldane22}
\dfrac{\epsilon}{|t_1|}=\pm 3\sqrt{3}\sin\varphi.
\end{equation}
In the original formulation, in which the dependence of $\varphi$ on $\alpha$ is derived by means of the Peierls substitution~\cite{haldane,shao2008}, the whole phase diagram is accessible. However, since the Peierls substitution is incorrect \cite{ibanez-azpiroz2014}, the possible values of $\varphi$ are actually limited to a finite range that depends on $s$, as discussed in section \ref{sec:results} (see e.g. Fig. \ref{fig:t0-t1-phi}(d)).
This is shown in Fig. \ref{fig:C-phi-M-t1}, where the accessible region for each value of $s$ is represented by the vertical (dashed) lines. Actually, only a small portion of the nominal phase diagram can be accessed (see the inset), as the maximum allowed values of $\varphi$ are much smaller than $\pi$. In the figure, the dots represent a non-trivial topological state with $C=\pm1$. The fact that almost all these points lie in between the black solid  
lines~\footnote{We note that only a few points located at relatively large values of $|\varphi|$ and $|\epsilon|/|t_{1}|$ lie outside the region defined by Eq. (\ref{eq:gaphaldane22}), see Fig. \ref{fig:C-phi-M-t1}.} proves that - in the allowed accessible region - the phase diagram of the microscopic Hamiltonian is well described by the  analytical expression of Eq. (\ref{eq:gaphaldane22}) for the Haldane model.

\begin{figure}[t]
\centerline{\includegraphics[width=1.00\columnwidth]{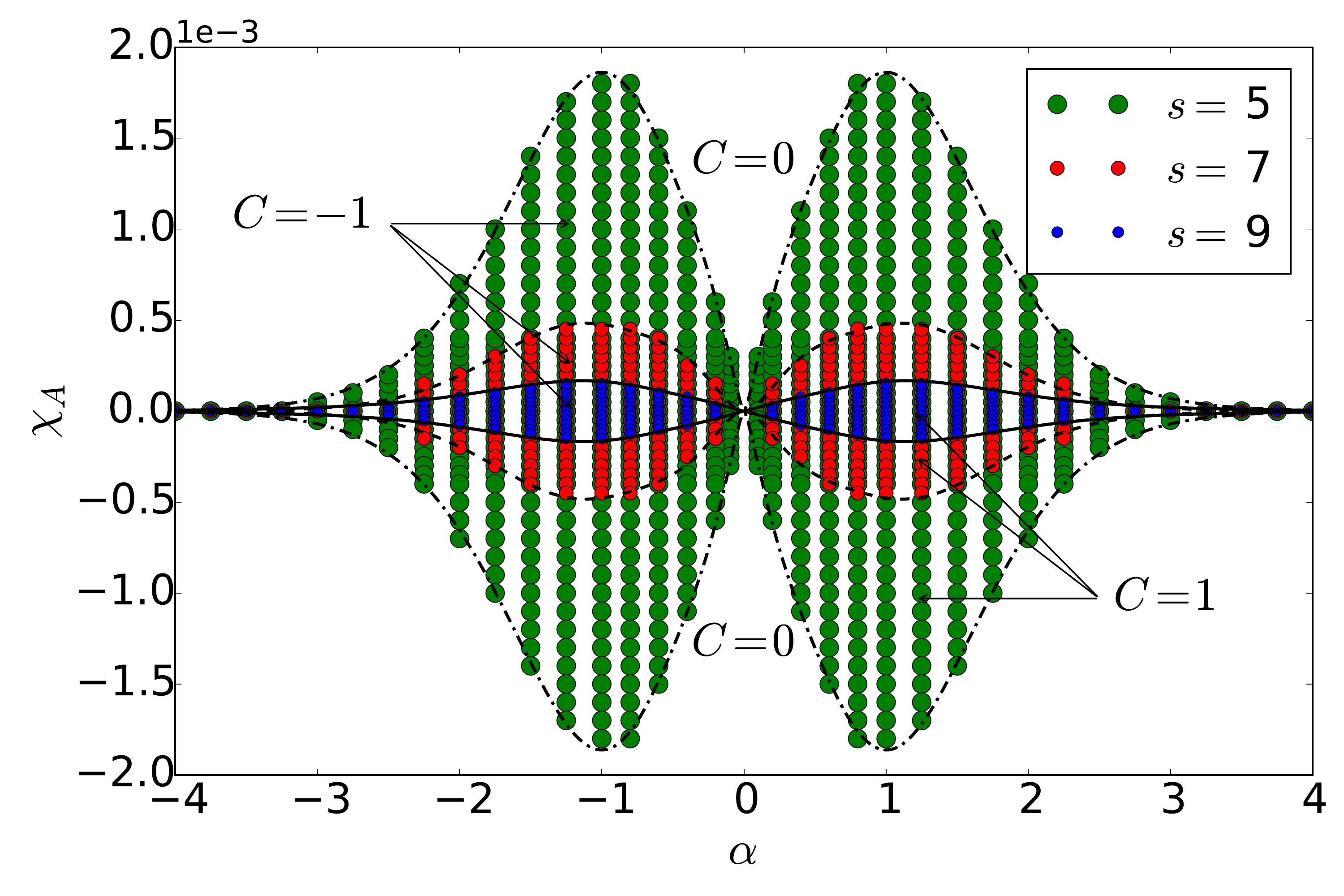}}
\caption{(Color online) Topological phase diagram of 
the continuous Hamiltonian in Eq. (\ref{eq:h0}), as a function of $\alpha$ and $\chi_{A}$, 
for three different values of the scalar potential amplitude $s$. 
The non-trivial topological state is indicated by big (green) dots for $s=5$, 
medium (red) dots for $s=7$ and small (blue) dots for $s=9$.
The black dashed lines represent a guide to the eye for the phase boundaries for each value of $s$.
}
\label{fig:C-chi-alpha}
\end{figure}
Owing to the above analysis, we suggest that a more appropriate way to draw 
the topological phase diagram is in terms of the physical parameters that
characterize the underlying continuous Hamiltonian, namely $\alpha,\chi_A$ and $s$.
This is shown in Fig. \ref{fig:C-chi-alpha}, where we plot the phase diagram in the 
$\alpha-\chi_A$ plane, for three different values of $s$. Importantly, the Fig. evidences 
that the topological insulating phase with $C\neq0$ shrinks dramatically as the system 
becomes more and more tight-binding (that is, by increasing $s$).  Notice that the sign 
of the Chern number in the topological insulator phase ($C=\pm1$) is consistent with the 
sign of $\alpha$, and independent on the sign of $\chi_{A}$. Notice also that the probability 
of finding the system in the topological insulator phase increases consistently by decreasing 
the value of $|\chi_{A}|$. 

\begin{figure}[t]
\centerline{\includegraphics[width=1.00\columnwidth]{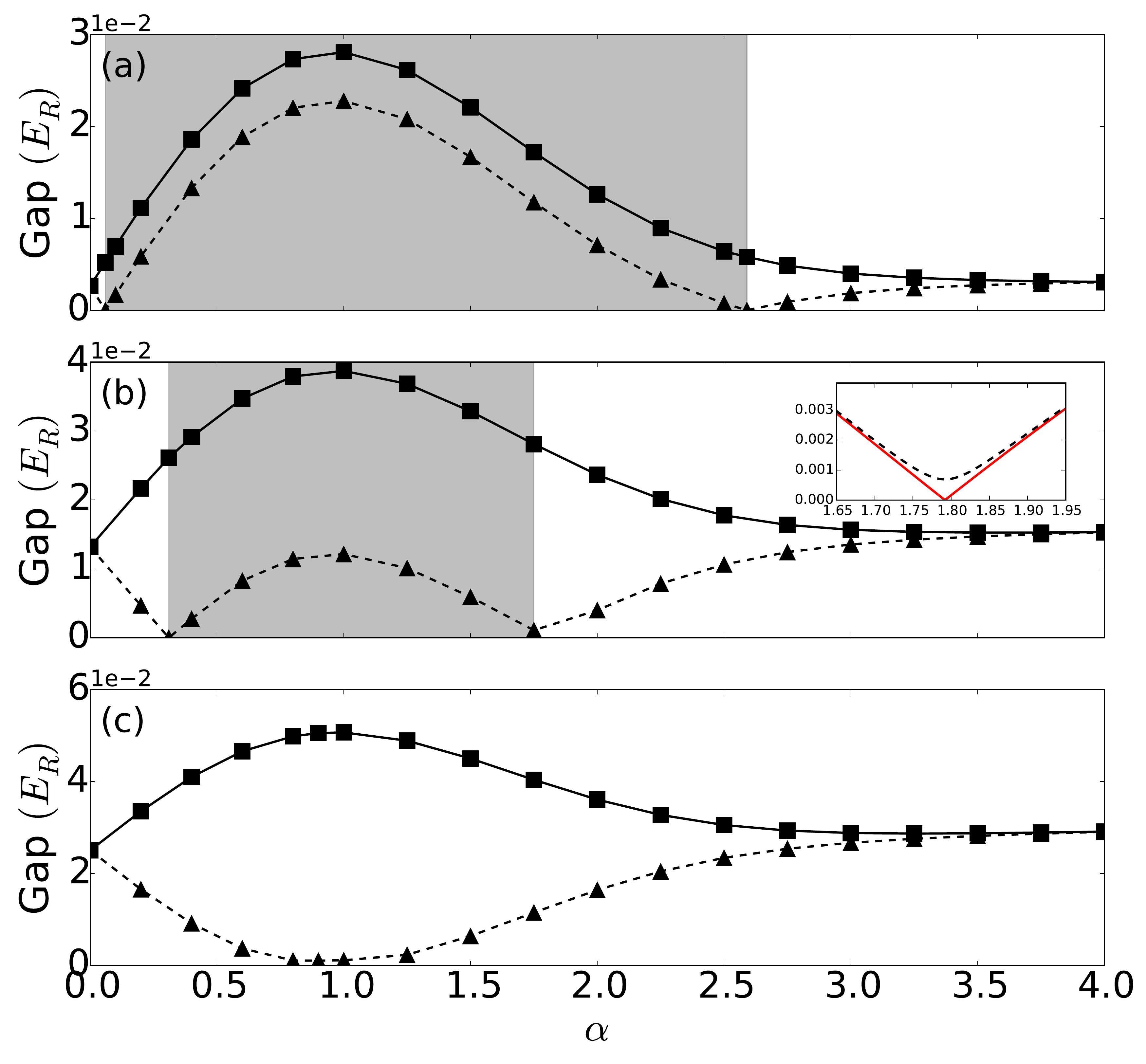}}
\caption{(Color online) behavior of the gaps $\delta_{+}$ (squares, solid line)
and $\delta_{-}$ (triangles, dashed lines) as a function of $\alpha$, for  $s=5$. The three panels correspond to  
$\chi_{A}=2\cdot10^{-4}$ (a), $10^{-3}$ (b), and $1.9\cdot10^{-3}$ (c).
The latter corresponds to the maximal value of $|\chi_{A}|$ for which the system can be in the topological insulating phase.
Grey shaded area corresponds to the region where the system is a topological insulator ($C=1$), whereas the white background identifies a normal insulating state ($C=0$).
The inset in panel (b) shows the behavior of the gap $\delta_{-}$ at $\bm{k}^{-}_{D}$ (dashed blue line) and at $\bm{\bar{k}}^{-}\simeq\bm{k}^{-}_{D}+1.68\cdot10^{-3}(1/2,\sqrt{3}/2)k_{L}$ (red continuous line), around the point $\alpha\approx1.8$.
}
\label{fig:gaps-Chern}
\end{figure}

As previously anticipated, the structure of the phase diagram is intimately 
connected to the behavior of the gaps at the Dirac points. 
This is illustrated in Fig. \ref{fig:gaps-Chern}, where we plot 
the gaps $\delta_{+}$ and $\delta_{-}$ as a function of $\alpha>0$
and three different values of  $\chi_{A}>0$ for fixed $s=5$. 
Noteworthy,
the gap closing does not take place exactly at $\bm{k}^{-}_{D}$
but in a close-by non-high-symmetry point. 
The origin of this feature has already been discussed in Sec. \ref{sec:haldane}.
For the specific value $s=5$ analyzed here, our calculations identify this point at
$\bm{\bar{k}}^{-}\simeq\bm{k}^{-}_{D}+1.68\cdot10^{-3}(1/2,\sqrt{3}/2)k_{L}$, as shown in the inset of Fig. \ref{fig:gaps-Chern}(b)~\footnote{We note that detecting such small shifts is 
not computationally demanding, since it is sufficient to do a thorough search of the 
neighborhood of the high symmetry point  $\bm{k}^{-}_{D}$, \textit{i.e.} not
of the full 1BZ. In addition, MLWFs need not be constructed for this step 
since only the eigenvalues are required.}.
We find that the gap closing point is slightly shifted for different values of $s$, but lies always
very close to $\bm{k}^{-}_{D}$.
In all cases, the deviation from $\bm{k}^{-}_{D}$ represents a minor correction, 
and can be safely ignored in the following
discussion.

Notably, Fig. \ref{fig:gaps-Chern} reveals that 
the gap has a maximum at $\alpha\simeq 1.0$ $k_{L}$, impliying that
the effect of the vector potential 
in opening the gap is limited, as expected from Eq. (\ref{eq:gaphaldane}).
It is also noteworthy that when $\chi_{A}$ is relatively small, as in Figs. \ref{fig:gaps-Chern}a and \ref{fig:gaps-Chern}b, the gap $\delta_{-}$ vanishes for two different values of $\alpha$ (the role of $\delta_{+}$ and $\delta_{-}$  is exchanged for $\alpha<0$) . 
In fact, owing to the non monotonic behavior of $\varphi$ as a function of $\alpha$, see Fig. \ref{fig:t0-t1-phi}d and Ref. \cite{ibanez-azpiroz2014}, there are two different values of $\alpha$ for which Eq. (\ref{eq:gaphaldane22}) can be satisfied 
(notice that the two values of $\varphi$ at 
the phase boundaries may be slightly different due to the fact that 
$t_1$ also depends on $\alpha$).
The intermediate region between these two values, which is 
represented by a grey shaded area  in the figures, corresponds to a topological non-trivial state 
($C=1$) where the effect of time-reversal symmetry breaking is stronger than inversion symmetry breaking.
As mentioned above, the smaller the value of $\chi_{A}$,
the larger the region with $C\neq0$ as a function of $\alpha$. 
By increasing  $\chi_{A}$, the topological insulating phase shrinks and eventually disappears,
as shown in Fig. \ref{fig:gaps-Chern}(c). 

\begin{figure}[t]
\centerline{\includegraphics[width=0.8\columnwidth]{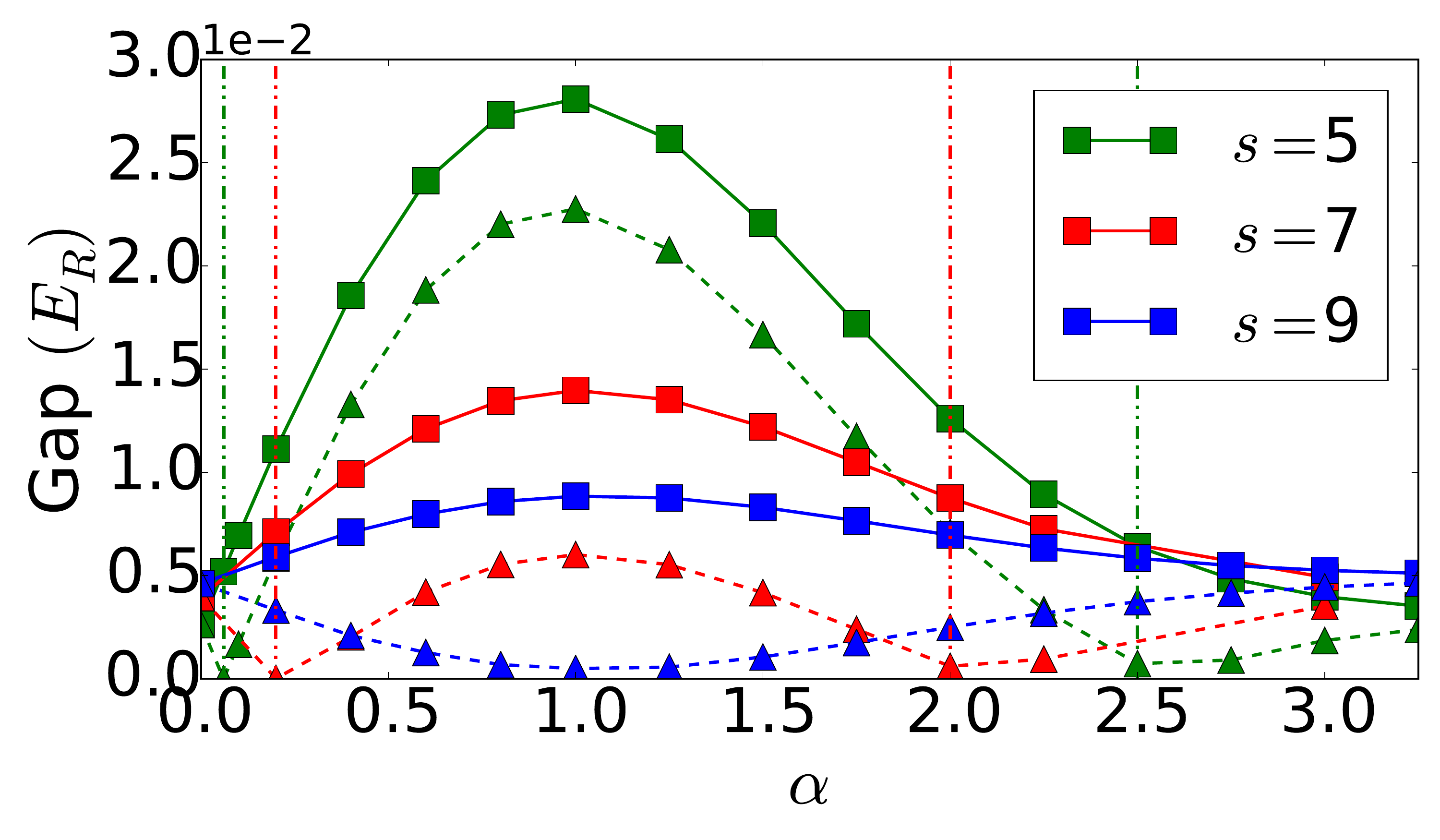}}
\caption{(Color online) Gap created at the Dirac points
by time-reversal symmetry breaking, for three different values
of the scalar potential amplitude $s$, fixed value $\chi_{A}=2\cdot10^{-4}$. 
Solid (squares) and dashed lines (triangles) denote $\delta_{+}$ and $\delta_{-}$, respectively.
The vertical dashed-dot-dot lines denote the gap closing points for the two lowest
values of $s$.
}
\label{fig:gaps-s}
\end{figure}

To conclude our analysis, let us discuss why the phase diagram of Fig. \ref{fig:C-chi-alpha} shrinks
as $s$ is increased. For such purpose, in Fig. \ref{fig:gaps-s} we illustrate the evolution of the gap
as a function of $\alpha$ and $s$ for fixed $\chi_{A}$. This
figure evidences that the maximum of the gap decreases as $s$ is increased; in other words,
the relative effect of time-reversal symmetry breaking decreases with increasing $s$.
As a consequence, even relatively low values of $\chi_{A}$ can avoid gap closing provided 
$s$ is large enough, as in the case of $s=9$ in Fig. \ref{fig:gaps-s}.
This, in turn, implies that the phase transition to the topological insulator
phase is restricted to smaller values of $\chi_{A}$ as $s$ is increased,
in agreement with the phase diagram of  Fig. \ref{fig:C-chi-alpha}.

\section{Conclusions}
\label{sec:conclusions}

In summary, we have presented an \textit{ab initio} analysis of a continuous Hamiltonian~\cite{shao2008}
that maps into the celebrated Haldane model~\cite{haldane}. 
The tunnelling coefficients of the tight-binding model have been computed by means of two independent methods,
one based on the maximally localized Wannier functions and the other on a closed set of analytical expressions in terms of 
the energy spectrum at selected high symmetry points in the BZ.
The two approaches present a remarkable agreement. In particular, we have shown that the gaps created either by inversion or time-reversal
symmetry breaking are very well described by the tight-binding model, which  
reproduces accurately the exact behavior.
In addition, we have calculated the topological phase diagram in terms of the 
physical parameters entering the
microscopic Hamiltonian, finding that only a small 
portion of the original phase diagram discussed by Haldane 
can be actually accessed within this model. Moreover, we have shown 
that the non-trivial topological phase with non-zero Chern number is 
suppressed as the system enters the deep tight-binding regime. 
We believe that, besides its conceptual implications, this work 
is relevant for a possible experimental
implementation of the Haldane model following the proposal in Ref. \cite{shao2008}.

\section{Acknowledgments}
We thank Y. Mokrousov, M. D. S. Dias, F. Guimar\~{a}es, P. Buhl, M. A. Vozmediano and J. Asboth for useful comments and discussions.
This work has been supported by the Universidad del Pais Vasco/Euskal Herriko Unibertsitatea under Program No. UFI
11/55, the Department of Education, Universities and Research of the Basque Government and the University of the Basque Country (IT756-13), the Ministerio de Econom\'ia y Competitividad through Grants No. FIS2013-48286-C2-1-P, No. FIS2013-48286-C2-2-P, No. FIS2010-19609-C02-00 and No. FIS2012-36673-C03-03, 
and the Basque Government through Grant No. IT-472-10. JIA. would like
to acknowledge support from the Helmholtz Gemeinschaft Deutscher-Young Investigators Group Program No. VH-NG-
717 (Functional Nanoscale Structure and Probe Simulation Laboratory)
and from the Impuls und Vernetzungsfonds der Helmholtz-Gemeinschaft Postdoc Programme.

\appendix

\section{Spread of the MLWFs}
\label{appendix:spread}

\begin{figure}[t]
\centerline{\includegraphics[width=0.8\columnwidth]{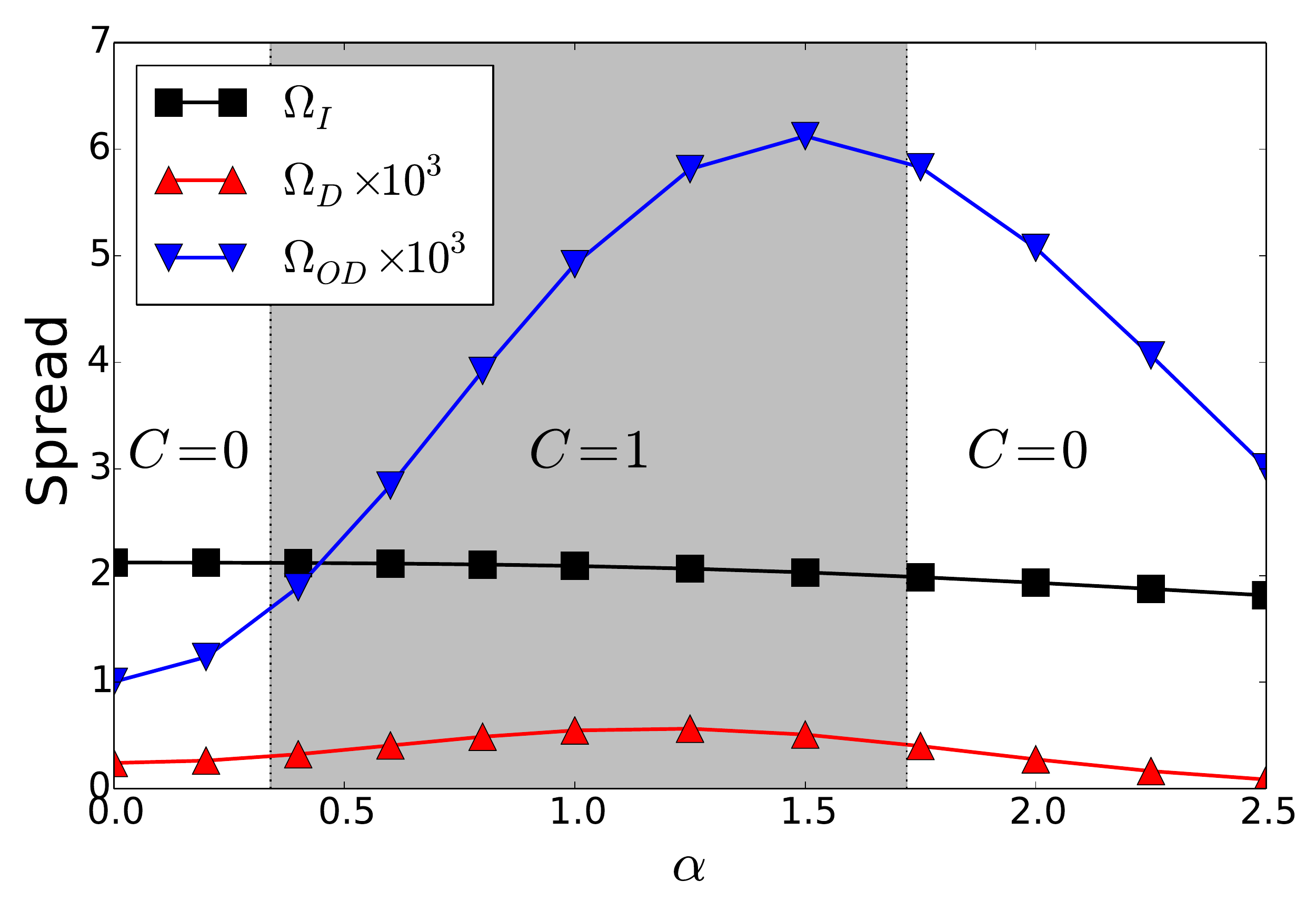}}
\caption{(Color online). Spread of the MLWFs as a function of
$\alpha$ for fixed values $s=5$, $\chi_{A}=1\cdot10^{-3}$.
The spread is decomposed into its gauge-invariant ($\Omega_{I}$), 
band diagonal ($\Omega_{D}$) and band off-diagonal ($\Omega_{OD}$)
terms. Note the $10^{3}$ factor in the case of $\Omega_{D}$ and $\Omega_{OD}$.
}
\label{fig:spread}
\end{figure}

Here we analyze the properties of the 
spread functional of the MLWFs,
$\Omega=\sum_{\nu}\left[\langle \bm{r}^2\rangle_{\nu}-\langle \bm{r}\rangle_{\nu}^{2}\right]$
~\cite{marzari1997}, 
as the amplitude $\alpha$ of the vector potential is varied 
and the system crosses the topological phase boundary.
Marzari and Vanderbilt showed that this functional can be divided into three parts,
namely $\Omega=\Omega_{I}+\Omega_{D}+\Omega_{OD}$~\cite{marzari1997}.  
The term $\Omega_{I}$ is gauge-invariant (namely, independent 
of the choice of the unitary transformations $U_{\nu\nu'}(\bm{k})$ in Eq. (\ref{eq:mlwfs})), whereas the
diagonal term $\Omega_{D}$ and the off-diagonal term $\Omega_{OD}$ do depend on the gauge choice.
In Fig. \ref{fig:spread} we show the behavior of the 
three terms of the spread as a function of $\alpha$, for fixed
values of $s$ and $\chi_{A}$. Here, the non-trivial
topological phase is indicated by the grey shaded area. 
All the components of the spread show a continuous behavior, even across the
boundary between the trivial and non-trivial topological states.
Then, it is interesting to note that, 
while the gauge-invariant term  $\Omega_{I}$ shows a monotonic decrease
as a function of $\alpha$,
the gauge-dependent terms $\Omega_{D}$ and $\Omega_{OD}$ 
show a non monotonic behavior that is reminiscent
of what we observed for the gap (see Fig. \ref{fig:gaps-Chern}) and for the complex phase of the next-to-nearest tunnelling coefficient in Fig.  \ref{fig:t0-t1-phi}(d).

We notice that the smooth behavior of the spread  shown by our calculations differs from 
an earlier analysis of MLWFs in the context of the Haldane model
performed by Thonhauser and Vanderbilt~\cite{thonhauser2006}. There the authors 
found a breakdown of the usual procedures to build MLWFs 
as the system approaches the topological phase boundary, 
resulting in a divergence of the spread functional.
The fundamental difference between our approach and the one followed 
in Ref. \onlinecite{thonhauser2006} resides in the set of 
bands considered for the construction of the MLWFs. 
In fact, whereas our set includes both the valence and conduction bands, their approach
included only the valence band. This is a crucial difference, since 
the net Chern number of a single band in the topological phase 
is finite, therefore it becomes impossible to choose a 
smooth periodic $\bm{k}$-space gauge of the Bloch orbitals and the procedure for 
constructing the MLWFs fails.
In our case, in contrast, the net sum of the Chern numbers of the valence and conduction
bands remains null, hence there is no formal impediment for the construction of the MLWFs.

\section{Numerical calculation of the spectrum}
\label{app:spectrum}

Both the calculation of the exact Bloch spectrum of the continuous Hamiltonian of Eq. 
(\ref{eq:h0}) and the construction of 
the MLWFs require a standard Fourier decomposition that here is 
adapted to account for the presence of the vector potential.  
We express the eigenstates $\psi_{n\bm{k}}(\bm{r})$ of the Hamiltonian 
as
\begin{equation}
\label{eq:pw-expansion}
\psi_{n\bm{k}}(\bm{r}) = \sum_{\bm{G}} c_{n\bm{k}+\bm{G}}e^{i\bm{G}\cdot\bm{r}},
\end{equation}
with  $\bm{G}$ the reciprocal vectors and $c_{n\bm{k}+\bm{G}}$ the expansion coefficients. 
The vector potential acts as 
$\bm{A}(\bm{r})\cdot \bm{p}+\bm{p}\cdot\bm{A}(\bm{r})=-2i\hbar \bm{A}(\bm{r})\cdot \bm{\nabla}_{\bm{r}}$,
introducing a non-local term when acting upon an eigenstate $\psi_{n\bm{k}}(\bm{r})$:
\begin{equation}
\label{eq:gradient}
i\bm{A}(\bm{r})\cdot\bm{\nabla}_{\bm{r}}\psi_{n\bm{k}}(\bm{r})=
-\bm{A}(\bm{r})\cdot\sum_{\bm{G}}\bm{G}c_{n\bm{k}+\bm{G}}e^{i\bm{G}\cdot\bm{r}}.
\end{equation}
Numerically, we found that a large number of $\bm{G}$ vectors are needed in order to 
converge the above term  due to the presence of the gradient. 
In particular, the above term requires an energy cutoff of $50$ $E_{R}$, 
whereas the rest of the terms in the Hamiltonian are converged with $10$ $E_{R}$.

Finally, for extracting the tight-binding parameters using the formulas discussed in section \ref{subsec:assumptions-haldane}, we have used a direct diagonalization of $H_{0}$ in Eq. (\ref{eq:h0}) by means of a standard Fourier decomposition. In this case, the vector potential term in Eq. (\ref{eq:gradient}) is 
transformed into a non diagonal matrix in momentum space.

\section*{References}
\bibliography{biblio}

\end{document}